\documentclass[onecolumn]{pasj00}

\begin{document}
\SetRunningHead{M. Takizawa}
               {Evolution of Hard X-Ray Radiation from Clusters of Galaxies}

\title{Evolution of Hard X-Ray Radiation from Clusters of Galaxies:
       \\ Bremsstrahlung or Inverse Compton Scattering?}

\author{Motokazu \textsc{Takizawa}%
        }
\affil{Department of Physics, Yamagata University, 
       Kojirakawa-machi 1-4-12, Yamagata 990-8560}
\email{takizawa@sci.kj.yamagata-u.ac.jp}
\affil{Department of Astronomy, University of Virginia, 
       P.O. Box 3818, Charlottesville, VA 22903-0818, USA}


%


\maketitle

\begin{abstract}
We have calculated evolution of a non-thermal electron population
from super-thermal but weakly relativistic to highly
relativistic energy range in clusters of galaxies. We investigate evolution 
of hard X-ray radiation due to both bremsstrahlung and inverse Compton 
scattering of the cosmic microwave background photons. The bremsstrahlung 
component is more significant than the inverse Compton scattering one
when the momentum spectra of electron sources are 
steeper than $\sim P_e^{-3.0}$ and vice versa in the case of Coma, 
where $P_e$ is an electron momentum. The resultant hard X-ray 
spectra are flatter when the bremsstrahlung component is dominant.
When the spectral indices of the source term are in the intermediate range 
($ -2.5 \sim  -3.5$), too much extreme ultraviolet emission is produced.
Inverse Compton dominant models can reproduce Coma cluster results with 
reasonable injection rates, which are possible in cluster mergers 
and/or ambient gas accretion.
\end{abstract}

\section{INTRODUCTION}\label{s:intro}

Clusters of galaxies (CG) contain not only the 
thermal intracluster medium (ICM) but also non-thermal
high energy particles in intracluster space. One of their direct
evidences is existence of non-thermal synchrotron radio halos and
relics, which have been observed since 1970s (Wilson 1970;
Giovannini, Tordi, \& Feretti 1999). This
indicates that there are non-thermal electrons with energy of 
$\sim$  GeV in intracluster space. Although their origin is
still not clear, they are probably related with some active
phenomena in CG such as cluster mergers, AGNs, AGNs'
jets, and/or star burst in the member galaxies. 
Therefor, it is crucial to study 
non-thermal high energy particles in order to investigate active
phenomena in CG. Furthermore, such non-thermal electrons and
protons possibly have an effect on ICM dynamics (Rephaeli \& 
Silk 1995; Inoue \& Sasaki 2001) and  Sunyaev-Zel'dovich effect 
(Blasi, Olinto, \& Stebbins 2000; Ensslin \& Kaiser 2000).

On the other hand, non-thermal hard X-ray
radiation due to inverse Compton scattering (ICS) of cosmic
microwave background (CMB) photons by the same 
electron population is expected from such CG with radio
halos and relics (Rephaeli 1979). 
There have been many attempts to detect such high energy
emission (e.g., Rephaeli, Ulmer, \& Gruber 1994;
Henriksen 1999; Valinia et al. 1999).
However, because of the dominant thermal 
X-ray emission from ICM, it was very difficult to detect
non-thermal hard X-ray radiation. 

Recently, excess of hard X-ray radiation over thermal
emission is detected from Coma (Fusco-Femiano 
et al. 1999; Rephaeli, Gruber, \& Blanco 1999), A2256
(Fusco-Femiano et al. 2000), and HCG62 (Fukazawa et
al. 2001) though their origin is still unclear. 
If we assume all hard X-ray radiation is due to ICS of CMB, 
we can estimate intracluster magnetic field strength or
its lower limit by comparing hard X-ray flux or its upper limit 
with synchrotron radio flux, respectively 
(e.g., Fusco-Femiano et al. 1999; Henriksen 1999;
 Valinia et al. 1999). However, 
the magnetic field strength estimated through this method
tends to be weaker than that determined through Faraday rotation
measurement (Kim, Tribble \& Kronberg 1991;
Clarke, Kronberg,\& B\"{o}hringer 2001). 

One possible solution of this discrepancy
is that all or a significant part of hard X-ray emission is due to
non-thermal bremsstrahlung (NTB) from weakly relativistic
non-thermal electrons with energy of a few 10 or 100 keV
(e.g., Kaastra et al. 1999; Ensslin, Lieu, \& Biermann 1999; 
Kempner \& Sarazin 2000; Dogiel 2000; Blasi 2000).
However, NTB needs much more non-thermal electron energy than ICS 
to produce a given amount of hard X-ray. In addition, dominant cooling 
process of electrons relevant to NTB hard X-ray is not bremsstrahlung 
losses but Coulomb interaction with thermal ICM, which results in
energy transfer from the non-thermal electrons to the thermal ICM.
Petrosian (2001) showed that unusual
rapid heating will occur if Coma hard X-ray is due to NTB.
However, calculation of NTB spectra is still interesting because it will
possibly work well to explain future observations of other clusters.
At present, hard X-ray excess have been detected from only a few clusters 
of galaxies. Thus, it is not clear whether Coma cluster is a typical case
or not.

Whether ICS or NTB is dominant in hard X-ray depends
on a shape of electron momentum spectra. This depends on 
acceleration and cooling processes of electrons. For
instance, if turbulent acceleration in ICM is an only
effective acceleration process, it is difficult to
accelerate thermal electrons up to $\sim$GeV because the
acceleration time is too long for typical intracluster
conditions. Therefore, almost hard X-ray is due to NTB (Blasi 2000).
On the other hand, if we
consider shock acceleration in ICM (Takizawa \& Naito 2000), 
it is expected that accelerated electron momentum spectra
are power-law up to the critical energy where an acceleration timescale is
comparable to an ICS cooling timescale. For typical rich clusters,
this critical energy is a few TeV
(Loeb \& Waxman 2000; Totani \& Kitayama 2000). 
Thus, both ICS and NTB can contribute to hard X-ray
radiation.
When electron momentum spectra are single power-law,
Sarazin and Kempner (2000) shows that NTB is dominant in hard X-ray 
if the momentum spectral index is more than $\sim$ 2.7.

Although accelerated electron momentum spectra in shocks
are single power-law, total spectra in CG do not
have to have the same form because various cooling processes
can change the spectral shape. In a higher energy range, ICS
and synchrotron loss modify the spectral shape from that of
the source. In lower energy range, on the other hand, Coulomb
loss does. Therefor, we have to to follow time evolution of
electron momentum spectra to investigate hard X-ray emission
from CG. 

In this paper, we calculate time evolution of
electron momentum spectra for spherically symmetric models of CG and
investigate evolution of both NTB and ICS contributions in
hard X-ray. It is certain that the synchrotron radio spectrum of Coma 
is known to a much higher precision than its hard X-ray spectrum. 
However, synchrotron emissivity is proportional to the non-thermal 
electron density and the magnetic field energy density. Unfortunately,
we do not have enough information about their spatial distribution. 
On the other hand, ICS and NTB emissivity are proportional to the 
non-thermal electron density and, the CMB energy density and the 
thermal ICM density, respectively. We have enough
spatial information about both of them from radio and X-ray 
observations. Thus, we think calculating the NTB and ICS hard X-ray 
spectra is as useful as calculating synchrotron radio ones in order 
to consider physical property of non-thermal electrons in clusters of 
galaxies. The rest of this paper is organized as
follows. In \S \ref{s:models} we describe our models. In
\S \ref{s:results} we present the results. In \S
\ref{s:coma} we compare our models to Coma cluster
and discuss their implications. Finally, our conclusions are 
presented in \S \ref{s:conclusions}.

\section{MODELS}\label{s:models}

\subsection{Evolution of Electron Momentum Spectra}
To follow evolution of momentum spectra of a non-thermal 
electron population, we should solve a diffusion loss
equation (Sarazin 1999). According to Bohm diffusion
approximation, however, the spatial diffusion term is negligible in
typical intracluster conditions within a few Gyrs (Takizawa
\& Naito 2000). We consider an isotropic momentum distribution function.
It is not clear whether the electron pitch angle distribution is
randomized or not in typical intracluster conditions.
However, this isotropic assumption is clearly valid when we consider 
Inverse Compton, bremsstrahlung, and Coulomb losses because
Faraday rotation measurements toward clusters of galaxies suggest tangled
magnetic field in intracluster space. 
Chaotic field lines also give rise to isotropic emission independent of
the electron pitch angle distribution. Only synchrotron loss rate depends on
the pitch angle distribution but it does not dominate other loss processes
within our parameter range.
We consider spherically symmetric ICM for
simplicity. We will use the normalized electron
momentum defined by $p_e \equiv P_e/(m_e c)$. 
Let $N(p_e, r, t)dp_e$ be the number density of non-thermal
electrons with momenta from $p_e$ to $p_e + d p_e$
at the radius $r$. Thus, the equation which describes
time evolution of $N(p_e, r, t)$ is,
\begin{eqnarray}
   \frac{\partial N(p_e, r, t)}{\partial t} =
       \frac{\partial}{\partial p_e}[b(p_e, r)N(p_e, r, t)]
        + Q(p_e, r, t),
\end{eqnarray}
where, $b(p_e, r)$ is the rate of normalized momentum loss for
a single electron with a normalized momentum of $p_e$ at the 
radius $r$, and $Q(p_e, r, t)$ gives the rate of production
of new non-thermal electrons.

We consider ICS of the CMB photons, synchrotron loss, and
Coulomb loss to determine $b(p_e, r$). Bremsstrahlung loss
is neglected because it does not contribute significantly in
our parameter range. Thus, 
\begin{eqnarray}
    b(p_e, r) = - \biggr( \frac{dp_e}{dt} \biggl)_{\rm IC} 
             - \biggr( \frac{dp_e}{dt} \biggl)_{\rm sync}
         - \biggr( \frac{dp_e}{dt} \biggl)_{\rm Coul}.
\end{eqnarray}
Each component is numerically (see Longair 1997),
\begin{eqnarray}
  -\biggr(\frac{dp_e}{dt}\biggl)_{\rm IC} &=& 
                 4.32 \times 10^{-4} \frac{p_e^2}{p_e^2+1}
                 p_e^2 {\rm Gyr}^{-1}, \\
  -\biggr(\frac{dp_e}{dt}\biggl)_{\rm sync} &=& 4.10 \times
     10^{-5} \frac{p_e^2}{p_e^2+1}
     p_e^2 \biggr( \frac{B}{\mu G} \biggl)^2
     {\rm Gyr}^{-1}, \\
  - \biggr( \frac{dp_e}{dt} \biggl)_{\rm Coul} &=&
             3.79  \times 10
             \biggr( \frac{n_e}{10^{-3} {\rm cm}^{-3} } \biggl)
             \frac{p_e^2+1}{p_e^2}
             \biggr( 1.0 + \frac{\ln(\frac{p_e^2}{p_e^2+1})}{19}
              + \frac{\ln(\frac{\sqrt{p_e^2+1}}{n_e/{\rm
             cm}^{-3}})}{75}
             \biggl) {\rm Gyr}^{-1},
\end{eqnarray}	
where $B$ is the intracluster magnetic field strength, and 
$n_e$ is the electron density of thermal ICM.
We consider CMB photons at z=0 as soft photon density calculating 
inverse Compton loss rate. We use 2.7 K as CMB temperature.

\begin{figure}
  \begin{center}
    \FigureFile(80mm,80mm){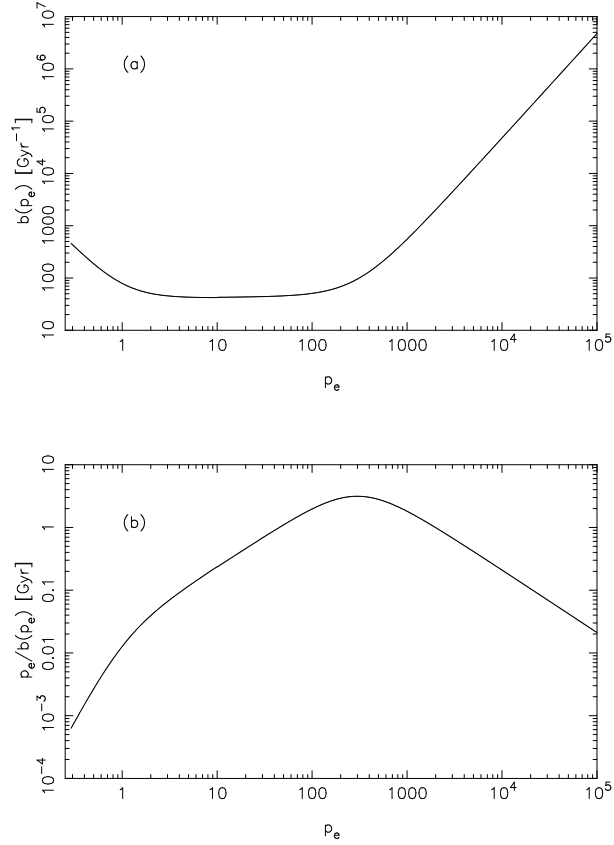}
  \end{center}
  \caption{(a)An electron momentum loss rate $b(p_e)$ as a function of $p_e$.
           (b)Electron cooling time defined by $p_e/b(p_e)$. In both cases 
              we assume $n_e = 10^{-3}$cm$^{-1}$ and $B = 1 \mu$G.
              ICS loss is dominant for a higher energy region 
              ($p_e \gg 100$), where $b(p_e) \propto p_e^2$ and 
              $p_e/b(p_e) \propto p_e^{-1}$. The loss rate is well 
              approximated by Coulomb loss with an extremely relativistic 
              limit in a middle region $(p_e \sim 10)$, where $b(p_e)$ is 
              almost constant and $p_e/b(p_e) \propto p_e$. 
              Transrelativistic effect appears in a lower energy part 
              ($p_e \sim 1$), where $b(p_e) \propto (p_e^2+1)/p_e^2$ and 
               $p_e/b(p_e) \propto p_e^3/(p_e^2+1)$ approximately.}
  \label{fig:bandtc}
\end{figure}

The loss rate $b(p_e)$ and cooling time defined by $p_e/b(p_e)$ 
are shown
in Figure \ref{fig:bandtc} as a function of $p_e$ when
$n_e = 10^{-3}$cm$^{-1}$ and $B = 1 \mu$G. ICS loss is
dominant for a higher energy region ($p_e \gg 100$), where
$b(p_e) \propto p_e^2$ and $p_e/b(p_e) \propto p_e^{-1}$.
The loss rate is well approximated by Coulomb loss with 
an extremely relativistic limit in a middle region 
$(p_e \sim 10)$, where $b(p_e)$ is almost constant and
$p_e/b(p_e) \propto p_e$. Transrelativistic effect appears
in a lower energy part ($p_e \sim 1$), where 
$b(p_e) \propto (p_e^2+1)/p_e^2$ and 
$p_e/b(p_e) \propto p_e^3/(p_e^2+1)$ approximately.

We assume that $Q(p_e,t) \propto p_e^{-\mu}$, which is
expected from the standard theory of shock acceleration.
We neglect the temporal change of the index $\mu$.

\subsection{Emission from a non-thermal electron population}

Let $l_{\epsilon}(r)$ be the volume emissivity of non-thermal hard
X-ray at energies from $\epsilon$ to $\epsilon + d \epsilon$ 
at the radius $r$.
We consider NTB and ICS of CMB photons from a non-thermal 
electron population,
\begin{eqnarray}
    l_{\epsilon}(r) = l_{\epsilon}^{\rm NTB}(r) +
                            l_{\epsilon}^{\rm ICS}(r).
\end{eqnarray}

For NTB components, the emission is given by the integral
(see Sarazin \& Kempner 2000),
\begin{eqnarray}
   l_{\epsilon}^{\rm NTB}(r) = \epsilon \int N(p_1,r)d p_1 v(p_1)
                 \sum_{Z} n_{Z}(r) 
                 \frac{d \sigma(p_1,\epsilon, Z)}{d \epsilon},
\end{eqnarray}
where, $v(p_1)= c p_1 /(p_1^2+1)^{1/2}$ is the velocity of
an electron with normalized electron momentum $p_1$, $Z$ is the
charge of the various thermal particles in ICM, $n_Z(r)$ is
the number densities of these thermal particles at $r$, and 
$[d \sigma(p_1, \epsilon, Z)/d \epsilon] d\epsilon$ is the 
bremsstrahlung cross section for emitting photons with
energies from $\epsilon$ to $\epsilon + d \epsilon$ for
the collision between an electron with normalized momentum
$p_1$ and an ion with electric charge $Ze$.

To calculate the cross sections, we use the
Born-approximation formulas of completely unscreened nuclei
(Koch \& Motz 1959). We use the extreme-relativistic limit
of Bethe-Heitler formula for the extremely  relativistic 
($p_e > 4.81$) electrons. On the other hand, we use the 
transrelativistic expressions of equations (2) and (4) in
Haug (1997) for weakly relativistic electrons ($p_e < 0.965 $) and 
moderately  relativistic ($0.965 < p_e < 4.81 $) electrons, respectively. 
The distortion of the electron wave function by the nuclear Coulomb
field is considered by multiplying the Elwert factor.

For ICS components, the emission is 
(see Blumenthal \& Gould 1970),
\begin{eqnarray}
 l_{\epsilon}^{\rm ICS} (r) = \frac{12 \pi \sigma_T}{h}
      \int N(\gamma,r) d \gamma 
      \int_0^1 J\biggr( \frac{\epsilon}{4 h \gamma^2 x} \biggl)
      F(x) dx,
\end{eqnarray}
where $\sigma_T$ is the Thomson cross section, $h$ is the Plank
constant and
\begin{eqnarray}
    F(x) \equiv 1 + x + 2 x \ln x - 2 x^2 .
\end{eqnarray}
$J(\nu)$ is the mean intensity at frequency $\nu$ of the
seed photon field. For the CMB, this is the black body
function
\begin{eqnarray}
  J(\nu) = \frac{2 h \nu^3}{c^2} 
           \frac{1}{\exp(h \nu/kT_{\rm CMB})-1}.
\end{eqnarray}

The luminosity at energies from $\epsilon$ to $\epsilon + d
\epsilon$ from a whole cluster $L_\epsilon$ is obtained by spatial
integration of the above-mentioned volume emissivity,
\begin{eqnarray}
 L_{\epsilon} = L_{\epsilon}^{\rm NTB} + L_{\epsilon}^{\rm ICS},
\end{eqnarray}
where, 
\begin{eqnarray}
 L_{\epsilon}^{\rm NTB} &=& \int l_{\epsilon}^{\rm NTB} 4 \pi r^2 dr, \\
 L_{\epsilon}^{\rm ICS} &=& \int l_{\epsilon}^{\rm ICS} 4 \pi r^2 dr.
\end{eqnarray}

\subsection{Model Clusters}

For the thermal ICM, we assume the isothermal beta model
(Cavaliere\&Fusco-Femiano 1978). The electron number density
of ICM at the radius $r$ is,
\begin{eqnarray}
  n_e(r) = n_{e,0} \biggr[ 1 + \biggr( \frac{r}{r_c} \biggl)^2
              \biggl]^{-\frac{3}{2}\beta}, 
\end{eqnarray}
and the temperature $T$ is constant. In order to compare our results
with Coma cluster in \S \ref{s:coma}, the parameters of Coma
are used. We set $n_0 = 3.12 \times 10^{-3} {\rm cm}^{-3}$, 
$r_c = 0.386 {\rm Mpc}$, and $\beta=0.705$ (Mohr, Mathiesen \&
Evrard 1999) and  $kT = 8.5 {\rm keV}$ (Fusco-Femiano et
al. 1999). We assume that $n_e(r) =0.0$ at $r>2$Mpc. This
outer boundary is roughly equal to that of the observed thermal
X-ray emission. As a result, the total number of thermal
electrons is $2.22 \times 10^{43}$ and the total thermal
energy of electrons is $4.52 \times 10^{63}$ erg.
We assume $B=1\mu$G. Evolution of non-thermal electrons 
hardly depends on the magnetic field strength unless
magnetic field energy density becomes comparable to or more than
CMB one.

For the non-thermal electron populations, we assume that 
constant injection for a fiducial period $t_{\rm inj}$ and that
momentum dependence of the injection rate is power-law.
We assume that this single power-law form spectrum is valid between
$p_{e, {\rm min}} < p_e < p_{e, {\rm max}}$, where
$p_{e, {\rm min}}$ is taken to be the normalized momentum
corresponding to $E_e = 3 kT$ and $p_{e, {\rm max}}$ is
assumed to be $10^7$, which is comparable to the critical
energy of typical rich clusters up to which shock accelerated
electron momentum spectra are power-low (Loeb \&
Waxman 2000; Totani \& Kitayama 2000).
Thus, $Q(p_e,r,t)$ is,
\begin{eqnarray}
    Q(p_e, r, t) = \left\{
                   \begin{array}{@{\,}ll}
                    Q_0(r) p_e^{-\mu} & (0 \le t \le t_{\rm inj}) \\
                    0                & (t_{\rm inj} < t),
                   \end{array}
                  \right.
\end{eqnarray}
where $Q_0$  has a dimension of number density per time.
Some candidates are proposed for the origin of a non-thermal
population. As for a electron component, however,
shocks in ICM caused by cluster merger is believed to be 
the most promising because they have enough and more energy 
than other candidates such as AGNs or normal galaxies.
Furthermore, it is difficult to explain the typical spatial size of
radio halos and relics for point sources like AGNs and
galaxies because diffusion length within ICS
cooling time is too short. In the
present model, thus, we set $t_{\rm inj} = 1$Gyr, which is roughly
equal to the duration of shocks in cluster mergers (Takizawa
1999, 2000). 

Spatial distribution of electron injection is not clear because
it depends on detailed parameters of cluster mergers. Thus we set
fairly simple assumption. We assume it is proportional to local
electron thermal energy. Total local injected energy of the non-thermal 
population is assumed to be 5\% of the total local energy of thermal 
electrons. Therefor, normalization of source term $Q_0(r)$ is,
\begin{eqnarray}
 Q_0(r) = 0.05 \times \biggr( \frac{n_e(r)}{t_{\rm inj}} \biggl)
          \biggr( \frac{3kT}{2m_e c^2} \biggl)
          \biggr( \int^{p_{e,{\rm max}}}_{p_{e,{\rm min}}}
           p_e^{-\mu} (\sqrt{p_e^2+1}-1)dp_e \biggl)^{-1}.
\end{eqnarray}
According to the standard theory of first-order Fermi acceleration, 
it is expected that exponents of momentum spectra $\mu$ are $2$ in
non-relativistic strong shocks (see Jones \& Ellison 1991). 
However, fairly low Mach number shocks are
expected in cluster mergers because the ICM of the pre-merger
clusters is already hot (Takizawa 1999, 2000; Miniati et al. 2000). 
In fact, some X-ray observations support
this idea (e.g., Markevitch, Sarazin \& Vikhlinin 1999;
Kikuchi et al. 2000). Thus we consider the cases
of $\mu = 2.0, 2.5, 3.0, 3.5$, and $4.0$

\section{RESULTS}\label{s:results}

\subsection{Evolution of Electron Momentum Spectra}

\begin{figure}
  \begin{center}
    \FigureFile(80mm,80mm){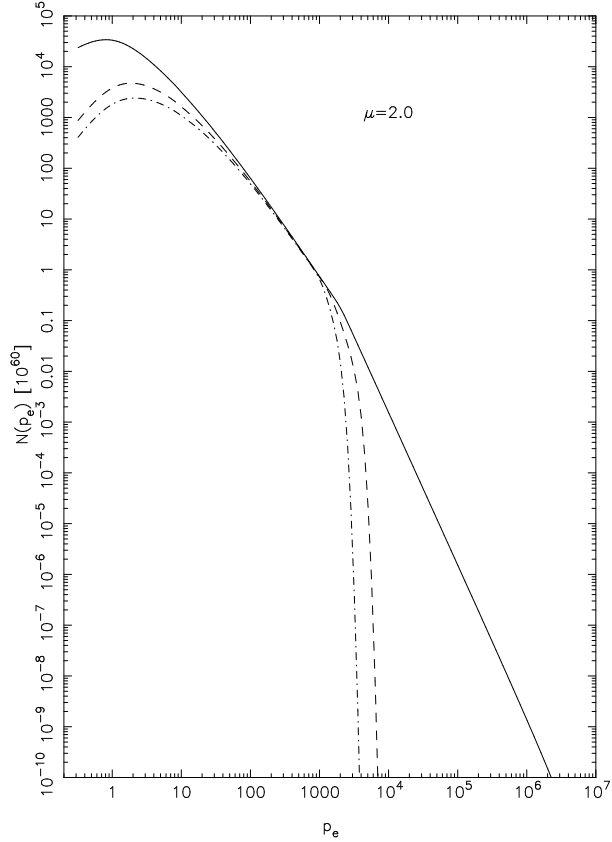}
  \end{center}
  \caption{Evolution of a total electron momentum spectrum in the
           $\mu = 2.0$ model. The different lines correspond to different 
           epochs: $t=1.0$ Gyr, solid line; $t=1.5$ Gyr, dashed line; 
           $t=2.0$ Gyr, dot-dashed line.
           While fresh non-thermal electrons are provided
          ($0<t<1$Gyr), the spectra change from the original single 
          power-law form into a steady state 
          solution. In the higher energy range the spectra change 
          from $N(p_e) \propto p_e^{-2}$ into $p_e^{-3}$ because of 
          ICS cooling. In the lower energy range, on the other hand, 
          the spectra change into $N(p_e) \propto p_e/(p_e^2+1)$
          because of Coulomb loss. After the injection stops ($t>1$Gyr), 
          inverse Compton cut-off appears and shifts towards lower energy 
          as time proceeds. In the lower energy range, the spectra change 
          into a steady state solution without a source term, 
          $N(p_e) \propto p_e^2/(p_e^2+1)$.}
  \label{fig:pspec}
\end{figure}

Figure \ref{fig:pspec} shows evolution of a total electron momentum
spectrum in the $\mu=2.0$ model. 
In all models electron momentum spectra evolve in a similar
way as follows. While fresh non-thermal electrons are provided
($0<t<1$Gyr), the spectra change from the original single power-law form 
($\propto p_e^{-\mu}$) into a steady state solution where
the injection balances with the cooling. In the higher energy range
the spectra change from $N(p_e) \propto p_e^{-\mu}$ into
$p_e^{-(\mu+1)}$ because ICS cooling is
dominant. In the lower energy range, on the other hand, 
the spectra change into $N(p_e) \propto p_e^{-\mu+3}/(p_e^2+1)$
because Coulomb loss is dominant. After the injection 
stops ($t>1$Gyr), inverse Compton cut-off appears and shifts towards 
lower energy as time proceeds. In the lower energy range, the
spectra change into a steady state solution without a source
term, $N(p_e) \propto p_e^2/(p_e^2+1)$. 

\subsection{Evolution of Hard X-ray radiation}

\begin{figure}
  \begin{center}
    \FigureFile(140mm,140mm){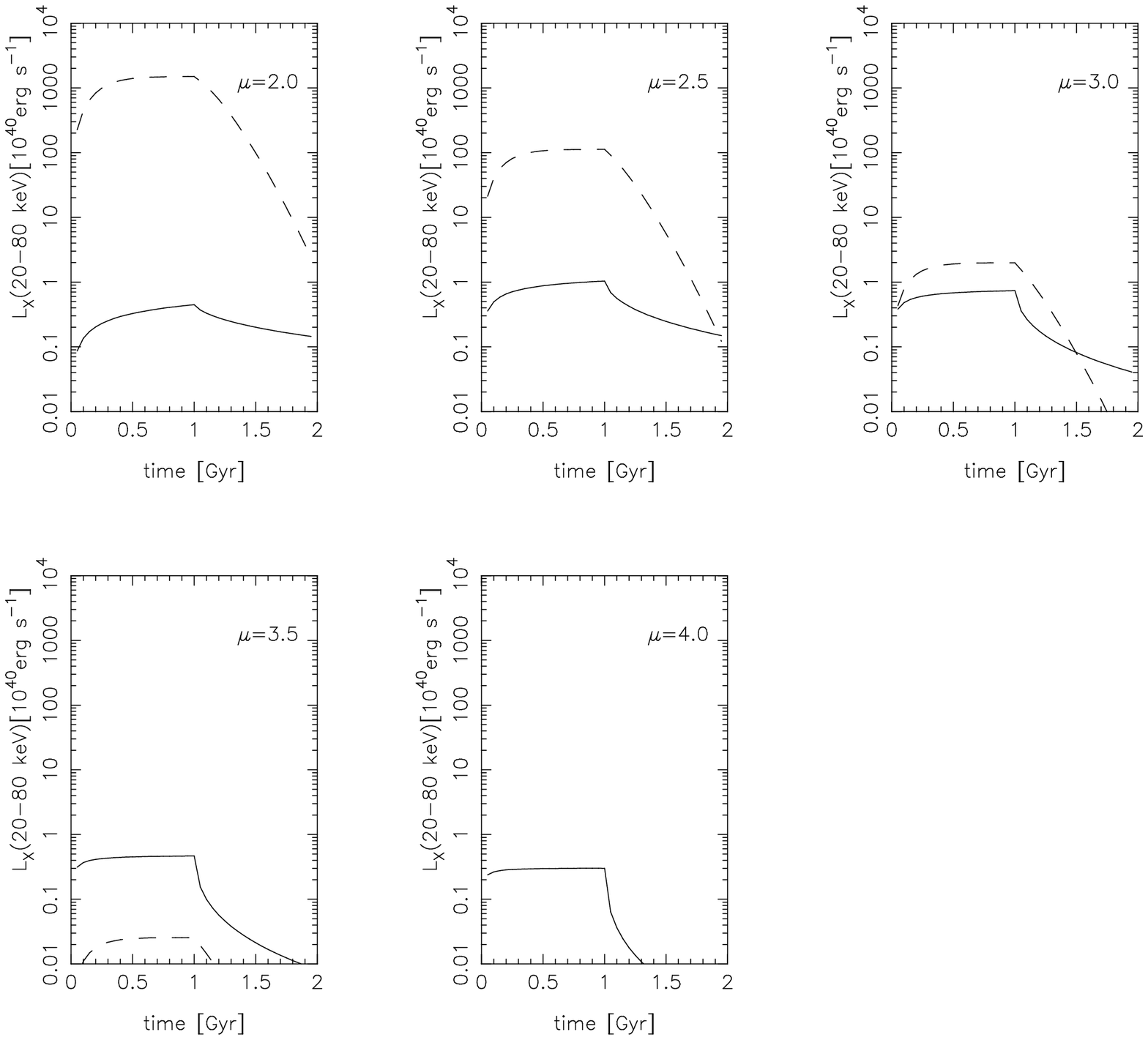}
  \end{center}
  \caption{Evolution of total hard X-ray (20-80 keV) luminosity of each 
           model. Solid and dashed lines show bremsstrahlung and inverse 
           Compton components, respectively. 
           The flatter source electron spectra result in more ICS and 
           less NTB emission. Both components 
           are comparable to each other in hard X-ray when $\mu \simeq 3$.
           }
  \label{fig:lumi}
\end{figure}

Figure \ref{fig:lumi} shows evolution of hard X-ray (20-80
keV) band total luminosity. Solid lines indicate NTB components
and dashed lines show ICS components. 
The flatter source electron spectra result in more ICS and less NTB
emission.
Both components 
are comparable to each other in hard X-ray when 
$\mu \simeq 3$. 
Sarazin \& Kempner (2000) shows that both components are
comparable to each other with $\mu \simeq 2.7$
when electron momentum spectra are single power-law.
However, note that we consider different situations 
In our models, electron sources
are power-law ($0<t<1$Gyr) or not active ($t>1$Gyr) and evolution of
electron spectra are fully calculated. In Sarazin \& Kempner
(2000), on the other hand, power-law electron spectra are
assumed for "Power-law" models, which correspond to only
a very early epoch ($t \sim 0$) in our models where cooling does not play
a significant role. They also consider "Cooling
Electrons" model, where spectra are assumed to be in a steady state
without any sources. This corresponds to a phase where source is not
active ($t>1$Gyr) in our models.
Our results are different than
theirs because electron momentum spectra are much
modified from a single power-law form in the source term 
because of the various cooling processes. In particular,
spectral flattening in lower energy owing to Coulomb cooling
is important.

Although total energy injected into a non-thermal electron
population is the same value in all models, 
the flatter the source electron spectra are,
the more luminous the total hard X-ray emission is. This is because
the electrons related to ICS components lose their
kinetic energy mainly through ICS loss while
the electrons related to NTB components lose their
kinetic energy mainly through not bremsstrahlung loss 
but Coulomb interaction with the thermal ICM. Thus, ICS
dominant models are much more efficient for energy transformation 
from non-thermal electrons into hard X-ray radiation. This
point is very important when the origin of non-thermal
electrons is discussed. 

After injection stops ($t>1$Gyr), both components of hard X-ray emission 
decrease in all models. However, timescales of decline 
are rather different between NTB and ICS components. In all
models ICS components decrease in the same way. The decline
timescale of ICS components is determined by ICS cooling
time of the relevant electrons, which is
commonly $\sim$ a Gyr for all models because energy of the
related electrons is $\sim$ GeV (see Fig.
\ref{fig:bandtc}b). However, this is not the
case for NTB components. The flatter the source spectra are,
the more slowly NTB components decrease. The reason of this
behavior is as follows. When the source spectra are flatter, 
equilibrium electron spectra are also flatter. Thus,
not only weakly relativistic electrons but also higher
energy ($\sim$ 100 MeV) electrons contribute to the hard X-ray 
through NTB significantly. While cooling time of weakly relativistic 
($\sim 10-100$ keV) electrons is very short 
($\sim 10^{6-7}$ yr, see Fig. \ref{fig:bandtc}b), that of
$\sim 100$ MeV electrons is
fairly long ($\sim$ a few Gyr). As a result, the contribution
from higher energy electrons survives for several Gyrs in
NTB hard X-ray. This is more prominent when the source
spectra are flatter. 

\begin{figure}
  \begin{center}
    \FigureFile(160mm,160mm){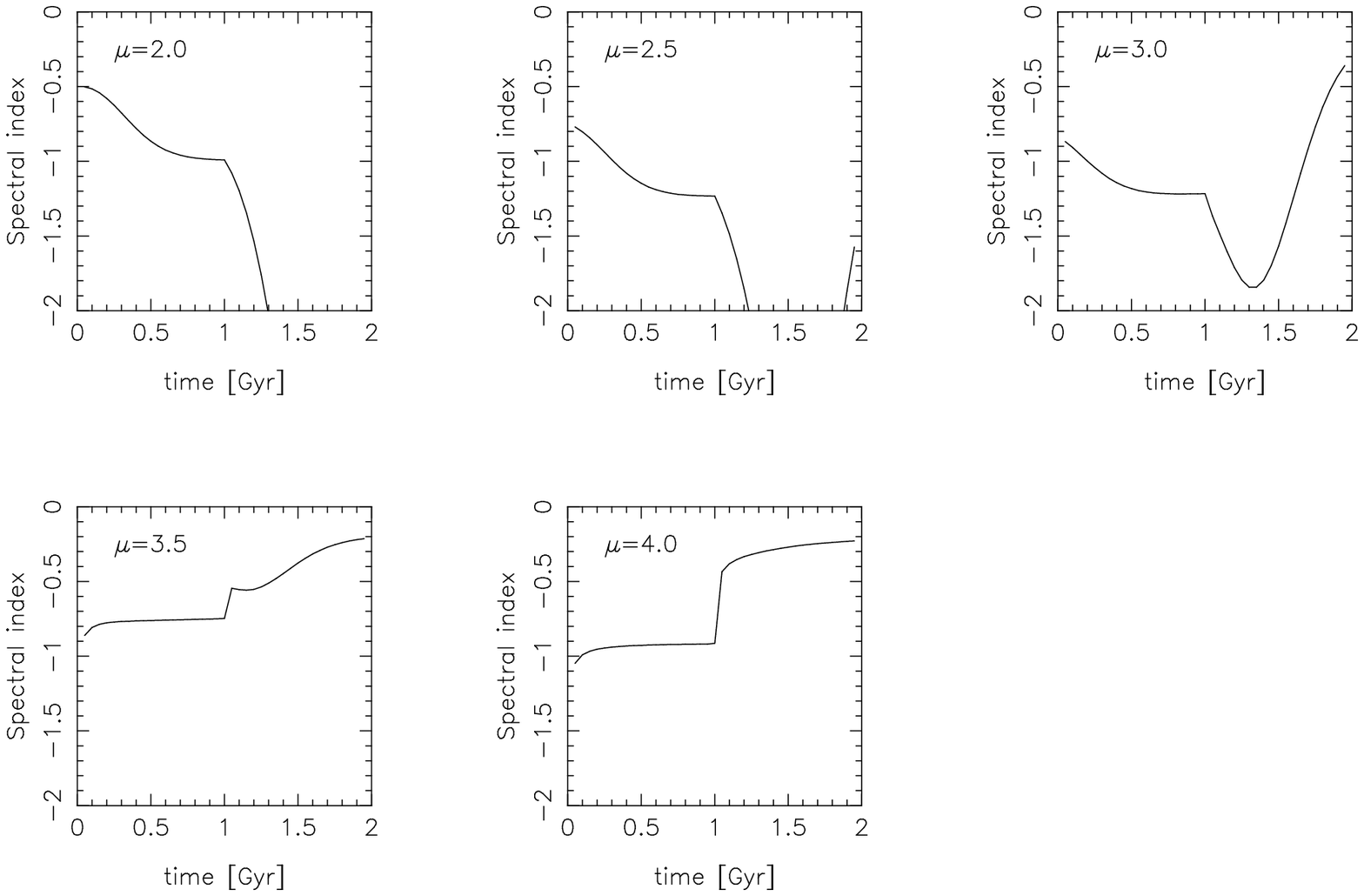}
  \end{center}
  \caption{Evolution of spectral indices in hard X-ray (20-80 keV) of 
           each model. The behavior depends on whether ICS or NTB 
           components are dominant. When ICS components are dominant 
           ($\mu < 3.0$), spectral indices $\alpha$ changes from 
           $-(\mu-1)/2$ into $-\mu/2$ while the source is active 
          ($0<t<1$Gyr). After the injection stops, the indices
          become much smaller values because the inverse Compton cut-off 
           enters a hard X-ray range. When NTB components are
           dominant ($\mu > 3.0$), the indices are $\sim -0.8$ 
           while sources are active. After the injection stops, 
           they become $\sim -0.15$.}
  \label{fig:sindex}
\end{figure}

Observed hard X-ray spectra are often parameterized by a single
power-law model. Thus, we fit our resultant total hard X-ray 
spectra into a single power-law model 
($L_{\epsilon} \propto \epsilon^{\alpha}$) to get a spectral
index $\alpha$. Figure \ref{fig:sindex} shows evolution of
the spectral indices of total hard X-ray (20-80 keV) spectra.
The behavior depends on whether ICS or NTB components are
dominant. When ICS components are dominant ($\mu < 3.0$),
spectral indices $\alpha$ changes from $-(\mu-1)/2$ into $-\mu/2$ while
the source is active ($0<t<1$Gyr). After the injection stops, the indices
become much smaller values because the inverse Compton cut-off 
enters a hard X-ray range. When NTB components are
dominant ($\mu > 3.0$), on the other hand, the indices evolve in a rather
different way. They become steady state values more
quickly. In this stage, the indices are $\sim -0.8$ within our
models. After the injection stops ($t>1$Gyr), they become
$\sim -0.15$, which is an approximation to a spectrum that varies as the
logarithm of the photon energy (see Sarazin \& Kempner 2000). 
In this way, NTB dominant models 
result in rather flat hard X-ray spectra.

\begin{figure}
  \begin{center}
    \FigureFile(150mm,150mm){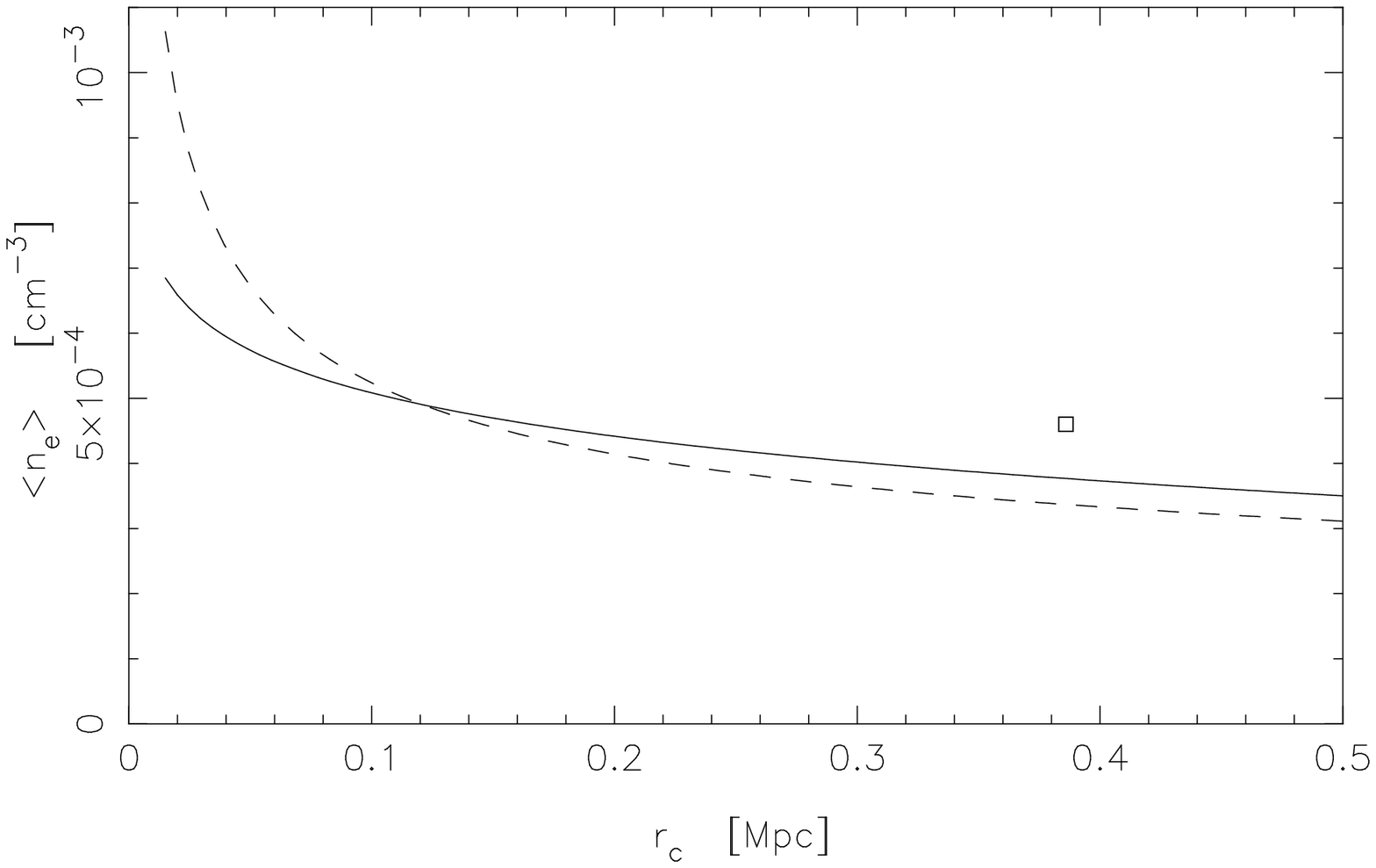}
  \end{center}
  \caption{Density-weighted mean density $<n_e>$ as a function of $r_c$ 
     when we consider a correlation between $r_c$ and $n_{e,0}$. Solid and dashed 
     lines indicate cases of $\beta=0.6$ and 0.7, respectively. A square point shows 
     a value of Coma cluster. $<n_e>$ can become about twice of that of Coma in 
     relatively high density clusters which follow the correlation. If we take 
     a dispersion in the correlation into account, $<n_e>$ can be nearly ten times 
     larger than that of Coma.}
  \label{fig:nbar}
\end{figure}

NTB flux is proportional to a volume integral of the square of the ICM 
density because we assume that injection of non-thermal 
electrons is proportional to the local ICM density. On the other hand, 
ICS flux is proportional to a volume integral of the ICM density itself. 
As a result, this ratio is proportional to density weighted mean 
density of ICM $<n_e>$ defined as follows,
\begin{eqnarray}
       <n_e> \equiv \frac{\int n_e(r)^2 r^2 dr}{\int n_e(r) r^2 dr}.
\end{eqnarray}	
X-ray observations shows that a core radius $r_c$ is correlated with 
central density $n_{e,0}$. Fujita \& Takahara (1999a, b) shows 
\begin{eqnarray}     
     \biggr( \frac{n_{e,0}}{{\rm cm}^{-3}} \biggl) 
     \simeq 6.4 \times 10^{-4} 
     \biggr( \frac{r_c}{{\rm Mpc}} \biggl)^{-1.3}
\end{eqnarray}
though a dispersion of a factor of three or four is seen in the data.
Using this correlation, we plot $<n_e>$ as a function of $r_c$ in 
figure (\ref{fig:nbar}). Solid and dashed lines indicate cases of $\beta=0.6$ 
and 0.7, respectively. A square point shows the Coma cluster.
This figure shows $<n_e>$ can become about twice of that of Coma
in relatively high density clusters which follow the correlation. 
If we take a dispersion in the correlation into account, $<n_e>$ 
can be nearly ten times larger
than that of Coma. Therefore, the flux ratio of NTB to ICS can be
about ten times higher than the values in figure 3 for relatively 
high density clusters. 

\section{COMPARISON WITH COMA CLUSTER}\label{s:coma}

In this section, we compare our models with the Coma cluster.
We compare a spectral shape of hard X-ray and a luminosity ratio
of hard X-ray to EUV of our models with Coma.
Then we seek an energy injection rate which can reproduce Coma 
for each model with different values of $\mu$ and discuss their 
implications. In Coma, a power-law model gives flux of 
$2.2 \times 10^{-11}$ ergs cm$^{-2}$ s$^{-1}$ in the 20-80 keV band
(Fusco-Femiano et al. 1999). This corresponds to 
luminosity of $5.1 \times 10^{43}$ erg s$^{-1}$ in the same
band. The spectral index of hard X-ray from Coma is poorly
determined. $0.3 \ge \alpha \ge -1.5$ is inferred at the 90\%
confidence level. In our parameter range of 
$\mu$ ($2.0 \le \mu \le 4.0$), thus, all models can
reproduce this fact while fresh non-thermal electrons are provided
(see a region of $0\le t \le 1$Gyr in Figure \ref{fig:sindex}).
However, it is difficult for ICS dominant models to reproduce
the above-mentioned range of $\alpha$ after the injection 
stops (see a region of $t>1$Gyr in Figure \ref{fig:sindex}).

\begin{figure}
  \begin{center}
    \FigureFile(160mm,160mm){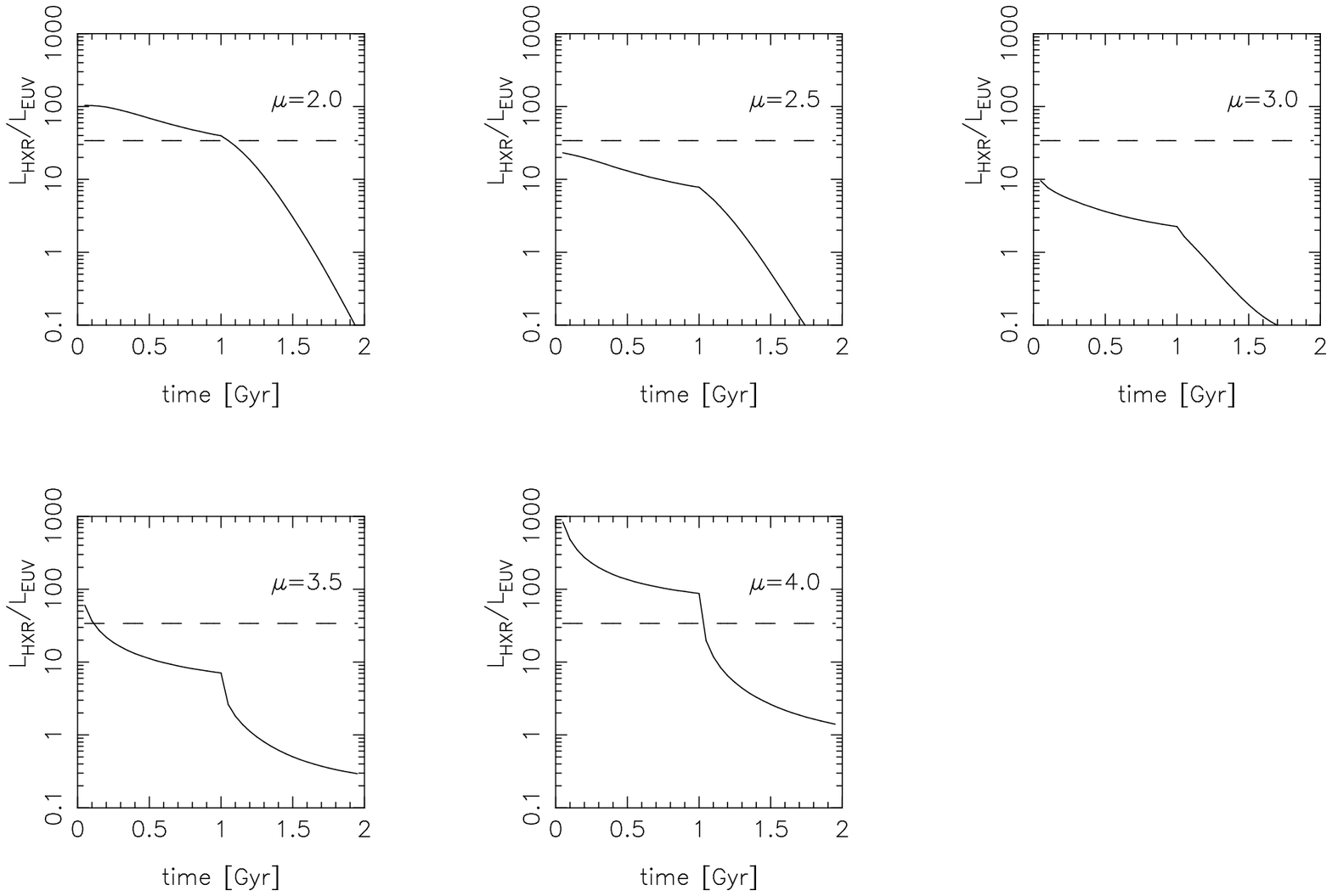}
  \end{center}
  \caption{Evolution of $L_{\rm HXR}/L_{\rm EUV}$ for each model
          (solid lines). Observed value of Coma derived from 
           Fusco-Femiano et al. (1999) and Bowyer et al. (1999) is
           shown by dashed lines. Note that $L_{\rm HXR}/L_{\rm EUV}$ 
           of our models must be more than the observed value. 
           It is clear that this constraint is not satisfied when 
           $\mu=2.5$ and $3.0$. When $\mu=3.5$, this can be satisfied 
           only in a very early epoch ($t<0.1$ Gyr). 
           On the other hand, models with $\mu=2.0$ and $4.0$
           satisfy the above-mentioned constraint in an active phase 
           ($0<t<1$Gyr).}
  \label{fig:hxeuvratio}
\end{figure}

Excess EUV radiation is also detected in Coma. The luminosity
ratio of hard X-ray to EUV can be used to constrain our models. In Coma, 
the excess EUV radiation is significantly detected only within $r < 0.64$Mpc
and a power-law model gives EUV luminosity of $1.5 \times 10^{42}$ erg
s$^{-1}$ in the 65-245 eV band (Bowyer, Bregh\"{o}fer \&
Korpela 1999). This gives a 
luminosity ratio $L_{\rm HXR}/L_{\rm EUV} = 34$. 
We calculate EUV luminosity due to ICS within 
$r< 0.64$ Mpc for each model and compare the ratios with
this value. 
Figure \ref{fig:hxeuvratio} shows evolution of
$L_{\rm HXR}/L_{\rm EUV}$ for each model (solid lines). An observed
value of Coma derived from Fusco-Femiano et al. (1999) and
Bowyer et al. (1999) is shown by dashed lines. 
Note that $L_{\rm HXR}/L_{\rm EUV}$ of our models must be
more than the observed value because ICS component from an old 
non-thermal electron population and/or thermal emission from
warm gas possibly contribute to the observed EUV emission, which
are not considered in our models. From figure
\ref{fig:hxeuvratio} it is clear that 
this constraint is not satisfied when $\mu=2.5$ and 
$3.0$. When $\mu=3.5$, this can be satisfied only in a very
early epoch ($t<0.1$ Gyr). In other words, these models 
emit too much EUV compared with hard X-ray.
On the other hand, models with $\mu=2.0$ and $4.0$
satisfy the above-mentioned constraint in an active phase ($0<t<1$Gyr).
The reason is as follows. When ICS is dominant in hard X-ray
($\mu<3$), the ratio of hard X-ray to EUV decreases as source
spectra are steeper. This is because electrons relevant to hard
X-ray have higher energy than those relevant to EUV. 
On the other hand, when NTB is dominant ($\mu>3$), that ratio increases
as source spectra are steeper because electrons relevant to hard
X-ray have lower energy than those relevant to EUV.
As a result, $\mu=2.0$ and $4.0$ models are favorable while 
$\mu =2.5, 3.0$, and $3.5$ models are not.

Although both ICS and NTB dominant models can be consistent 
with the observed hard X-ray spectral index of Coma, our NTB 
dominant models do not produce observed radio spectrum. For example, 
Deiss et al. (1997) shows the radio spectral index in the frequencies 
lower than 1.4 GHz is -1.16, which means $\mu=2.32$ in the electron 
source in an steady state. Therefor, another non-thermal electron 
component with flatter spectrum is necessary even if NTB is dominant 
in hard X-ray. 

Next, we seek a numerical
value of a total injected energy rate $\dot{E}_{\rm nth}$ 
which reproduce the Coma cluster hard X-ray
luminosity at $t=1$ Gyr. 
This is consistent with ICM temperature distribution
derived from X-ray observation (Watanabe et al. 1999) and 
some numerical calculations of cluster merger (Burns et
al. 1994; Ishizaka \& Mineshige 1996) which suggest that
Coma cluster is in major merger which started $\sim 0.5-1$
Gyr ago. Note that  $\dot{E}_{\rm nth}$ is not very sensitive to 
accurate determination of an observational epoch
because hard X-ray luminosity is nearly constant
in all our models in $0.5$Gyr$<t<1$Gyr.
Then, we calculate total injected energy to the 
non-thermal population in 1 Gyr, $E_{\rm tot, nth}$, and its
ratio to the total electron thermal energy, $E_{\rm tot, nth}/E_{\rm th}$.
The results for each model are shown in table \ref{tab:coma}.

From table 1, we find that NTB dominant models
require very high rates of energy injection to explain 
the results of Coma cluster.
This causes violent heating of thermal ICM 
unless duration of the acceleration process is very short
(Petrosian 2001). Thus, we conclude
that it is very difficult to consider NTB
components significantly contribute to hard X-ray emission
of Coma. 

On the other hand, ICS dominant models
are free from the difficulties mentioned above. The required
injection rate is $\sim 10^{46-47}$ erg s$^{-1}$. This value
can be realized by cluster mergers. Thus, ICS components
are more likely as an origin of Coma hard X-ray
in point of required energy. As many authors
pointed, however, the flux ratio of synchrotron radio to
hard X-ray gives us the volume averaged magnetic field
strength in this case,
which must coincide with that determined from
Faraday rotation toward individual polarized radio galaxies
unless spatial inhomogeneity plays an significant role.
(e.g., Fusco-Femiano et al. 1999; Ensslin, Lieu, \&
Biermann 1999; Kempner \& Sarazin 2000). 
We should look for any other causes of this discrepancy.
For example, we should consider the spatial structure of
magnetic field, inhomogeneity of magnetic field and
non-thermal electrons, or spatial anti-correlation of magnetic
field strength and non-thermal electron density, etc.
Selection bias and Galactic contamination are extensively discussed
by Goldshmidt \& Rephaeli (1993) as the possible origin of this 
discrepancy.

\section{CONCLUSIONS}\label{s:conclusions}

We calculate evolution of a non-thermal electron population 
in clusters of galaxies and investigate hard X-ray radiation
due to both NTB and ICS.
We calculate cases where the electron sources are
power-low in momentum ($\propto p_e^{-\mu}$). 
With the parameters of Coma, ICS components dominate NTB ones
in hard X-ray radiation when $\mu < 3.0$ and vice versa.
This dividing line may depend on thermal ICM density distribution.
NTB dominant models have fairly flat hard X-ray spectra.

We compare our models to the Coma cluster.
Luminosity ratio of hard X-ray to EUV is consistent
with Coma when $\mu=2.0$ and 4.0. When $\mu=3.5$,
it is consistent only in a very early epoch. 
However, NTB dominant models require very high
energy injection rates and cause unusual 
violent heating of thermal ICM unless 
duration of injection is very short (Petrosian 2001).
On the other hand, ICS dominant models require much lower injection
rates, which can be realized by cluster mergers. Thus,
we conclude that ICS is more likely as the origin of Coma
hard X-ray. 
In this case, however, the flux ratio of synchrotron radio to
ICS hard X-ray gives us the volume averaged magnetic field
strength. This value must coincide with that determined from
Faraday rotation toward individual polarized radio galaxies
unless spatial inhomogeneity plays an significant role.
Thus, we should look for other solutions 
for the apparent discrepancy between magnetic field strength 
derived from the above-mentioned two methods.
It may be related to spatial configurations
of cluster magnetic field.

Our conclusions suggest that the momentum spectrum of the electron
sources in Coma is fairly flat ($\mu < 2.5$). This is consistent
with radio observations. Deiss et al. (1997) shows the radio spectral
index in the frequencies lower than 1.4 GHz is $-1.16$, which
corresponds to $\mu = 2.32$ in the electron source term in
a steady state (Note that equilibrium electron spectra are
steeper than source spectra because of ICS cooling).
However, this implies fairly high Mach number shocks. According 
to the standard theory of shock acceleration, $\mu < 2.5$
means a Mach number more than 3.0. This value is somewhat
lager than estimated values ($\sim 2$) of merger shocks from 
X-ray observations of other clusters (e.g., Markevitch,
Sarazin, \& Vikhlinin 1999; Kikuchi et al. 2000). However,
estimated Mach numbers can be underestimated because 
observed temperature gradients can be underestimated owing to
contamination by foreground and/or background ICM
(Takizawa 2000, Shibata et al. 2001). Thus, it is possible
that such 'hidden' high Mach number shocks contribute to particle
acceleration in Coma. Indeed, merger shocks with a Mach number more 
than 3 do exist in numerical simulations of cosmological
structure formation (Miniati et al. 2000).
Another possibility is acceleration at
external accretion shocks, which can be very high Mach number
shocks because accreting gas is much colder than cluster
virial temperature (Miniati et al. 2000).

Another possibility is reacceleration in intracluster space
due to turbulent Alfv\'{e}n waves and/or other physical process
(Brunetti et al. 2001; Ohno, Takizawa, \& Shibata 2001).
In this case an electron spectrum can be much modified from an original
source form. As a result, situation effectively similar to our 
$\mu=2.0$ model possibly occurs though it depends on detailed acceleration
history of non-thermal electrons.

In general, it is believed that non-thermal energy is
less than thermal one in ICM. 
Only ICS dominant models with flat source spectra can
satisfy this situation among our models. Even in this case, however,
the obtained non-thermal to thermal energy ratio is fairly
high. For example, TeV gamma-ray observation
of SN1006 suggests this ratio is a few \% 
(Tanimori et al. 1998), whereas our $\mu=2.0$ model implies this
ratio is more than 10 \%.
However, note that total energy of non-thermal
electrons is sensitive to a spectral shape in lower
energy. We assume that source spectra are power-law 
in momentum down to $E_e = 3kT$. This is valid when
thermal electrons in ICM are the source of non-thermal electrons.
However, this is not the case when old non-thermal electrons
which leaked from AGNs and/or galaxies, and/or fossil
radio plasma (Ensslin \& Gopal-Krishna 2001) are the source.
Once electrons are accelerated up to $\sim 100$
MeV, they have a fairly long ($\sim$ a few Gyr) life time.
Thus, it is possible that lower boundary energy of single power-law 
is larger and that required energy is less.

\bigskip

I thank M. Henriksen and K. Nakazawa for fruitful discussion and 
S. Shibata for continuous encouragement.
I would also like to thank the anonymous referee for very useful comments.


 \begin{table}
  \caption{Comparison with the Coma cluster.}
  \label{tab:coma}
  \begin{center}
   \begin{tabular}{cccc}
    \hline \hline \\
  & $\log(\dot{E}_{\rm nth})$ & $\log(E_{\rm tot, nth})$ & $\log(E_{\rm tot, nth}/E_{\rm th})$ \\
Model & erg s$^{-1}$ & erg  &  \\
    \hline
$\mu=2.0$ & 46.39 & 62.89  & -0.77   \\
     2.5  & 47.51 & 64.01  &  0.35   \\
     3.0  & 49.13 & 65.63  &  1.97   \\
     3.5  & 49.87 & 66.37  &  2.71   \\
     4.0  & 50.08 & 66.58  &  2.93   \\
    \hline
   \end{tabular}
  \end{center}
 \end{table}%


\end{document}